\documentclass[useAMS,usenatbib]{mn2e}
\usepackage{graphicx}
\usepackage{fixltx2e} 
\usepackage{natbib}

\def\Vt{$V_{\rm t}$~} 
\def\EE{$E^{\rm ''}$}

\def\vsini{$v$ sin $i$~}
 
\def\tef{$T_{\rm eff}$~} 

\def\kms{km s$^{\rm -1}$~}

\begin{document}

\title[\tef, \vsini, \Vt and abundances in the Sun, HD1835 and HD10700]
{Effective temperatures, rotational velocities,
microturbulent velocities and abundances in the atmospheres of the Sun, 
HD1835 and HD10700}


\author[Ya. V. Pavlenko et al.]{Ya. V. Pavlenko$^{1,2}$\thanks
{E-mail: yp@mao.kiev.ua 
Based on observations made with the ESO telescopes at the La Silla Paranal 
observatory under programme ID's 076.C-0578(B) and 077.C-0192(A).}, 
J.S. Jenkins$^{2,3}$,
H.R.A.Jones$^{2}$, 
O.M. Ivanyuk$^{1}$, 
D.J. Pinfield$^{2}$\\
$^{1}$Main Astronomical Observatory, Academy of Sciences of Ukraine, 
Golosiiv Woods, Kyiv-127, 03680 Ukraine \\
$^{2}$Centre for Astrophysics Research, University of Hertfordshire,
College Lane, Hatfield, Hertfordshire AL10 9AB, UK\\
$^{3}$Departamento de Astronom\'ia, Universidad de Chile, 
Camino del Observatorio 1515, Las Condes, Santiago, Chile\\
}

\date{}
\pagerange{\pageref{firstpage}--\pageref{lastpage}} \pubyear{2002}

\maketitle

\label{firstpage}

\begin{abstract}

We describe our procedure to determine effective temperatures, 
rotational velocities, 
microturbulent velocities, and chemical abundances in the 
atmospheres of Sun-like stars.
We use independent determinations of iron abundances 
using the fits to the observed Fe I and Fe II 
atomic absorption lines. 
We choose the best solution 
from the fits to these spectral features for the 
model atmosphere that provides the best confidence in 
the determined log N(Fe), \Vt, and \vsini.Computations were done in
the framework of LTE. Blending effects were accounted for explicitly.
First, we compute the abundance of iron for a set of adopted microturbulent velocities. 
In some cases, a few points of log N(Fe I) = log N(Fe II) can be found. 
To determine the most self-consistent effective temperature and microturbulent 
velocity in any star's atmosphere, we
used an additional constraint where we minimise the dependence of the 
derived abundances of Fe I and Fe II on the excitation potential of the 
corresponding lines.  Using this procedure 
we analyse the spectra of the Sun and two well known solar 
type stars, HD1835 and HD10700 to determine their 
abundances, microturbulent velocity and rotational velocity.  
Our approach allows us to determine self-consistent values for the 
effective temperatures, abundances, \Vt and \vsini.
For the Sun  we obtain the best agreement for a model atmosphere of 
\tef/log g/[Fe/H] = 5777/4.44/0.0, iron abundances and 
microturbulent velocities of log N(Fe) =4.44, \Vt = 0.75 km/s, 
for the Fe I lines,
and log N(Fe) = -4.47, \Vt = 1.5 km/s for the Fe II lines.
Furthermore, abundances of other elements obtained from the fits of their 
absorption features agree well enough  
($\pm$ 0.1 dex) with the known values for the Sun. We determined a rotational velocity of 
\vsini = 1.6 $\pm$ 0.3 km/s for the spectrum of the Sun as a star. 
For HD1835 the 
self-consistent solution for Fe I and Fe II lines log N(Fe)=+0.2 was 
obtained with a model atmosphere of 5807/4.47/+0.2 and
microturbulent velocity \Vt = 0.75 km/s, and leads to \vsini = 7.2 $\pm$ 0.5 km/s. 
For HD10700 the self-consistent solution log N(Fe) = -4.93 was obtained using a 
model atmosphere of 5383/4.59/-0.6
and microturbulent velocity \Vt = 0.5 km/s.  The Fe I and Fe II lines give rise to 
a \vsini = 2.4 $\pm$ 0.4 km/s. 
Using the \tef found from the ionisation equilibrium parameters 
for all three stars, we found abundances of a number of other elements: Ti, Ni, Ca, Si, Cr.  
We show that uncertainties in the adopted values of \tef of 100 K
and \Vt of 0.5 km/s change the abundances of elements up to 0.1 and 0.2 dex
respectively. Galactic abundances variations can generally be larger than this
measurement precision and therefore we can study abundance variations
throughout the Galaxy.

\end{abstract}

\section{Introduction}

In the past the majority of research conducted on
solar-type stars was focused on accurately constraining their physical 
parameters, such as effective temperature, 
bolometric magnitude, radius, metallicity and color indices using a 
variety of different photometric and 
spectroscopic techniques (e.g. Glushneva et al. 2002). However, with 
the advance of technology, high 
resolution echelle spectrographs and more observing time, the 
field is pushing to even higher 
precision measurements for a larger database of atomic and molecular 
species in stellar atmospheres (e.g. 
Valenti \& Fischer 2005; Neves et al. 2009; Mashonkina et al. 2011)
and more precise physical parameters from the latest evolutionary 
models (Lopez-Santiago et al. 2010;
Ghezzi et al. 2010; Takeda et al. 2010;  Tabernero et al. 2011). 
One of the main reasons for 
the growing interest into solar-type 
stars is that the metal content of planet-hosting stars is an important 
ingredient that seems to affect the formation and evolution of planetary 
systems  (see Israelian 2010; Santos et al. 2011; Sousa et al. 2012, and 
reference therein). 

Valenti \& Fischer have published an extended uniform catalog of stellar 
properties for 1040 nearby F, G, and K 
stars that were observed by the  Keck, Lick, and AAT planet search programs. 
Accurate stellar abundances from 
spectral synthesis fitting requires the determination of reliable physical 
parameters, namely the effective temperature, 
surface gravity, microturbulent velocity, rotational velocity and 
atomic abundances (see Jenkins et 
al. 2008; Jenkins et al. 2009a).  Studying stars with spectral types similar 
to the Sun but with other physical differences can help to test new analysis 
techniques.  

HD1835 is a variable and magnetically active G3V star (see 
Jenkins et al. 2006, Mart\'{i}nez-Arn\'{a}iz, 2010) 
In Table \ref{_abu1835} we show the iron 
abundances studied by different authors for HD1835.  When we look at past 
studies there is a fairly large spread in 
metallicities for this star, however, 
in recent times there is more convergence 
towards the star having super-solar metallicity.  Only 
two studies found this star to have sub-solar metallicity, the last back 
in 1994.

 \begin{table}
 \centering
 \caption{\label{_abu1835} Temperatures, gravities and abundances used in 
studies of HD1835.}
 \begin{tabular}{ccccl}
 \hline
 \noalign{\smallskip}

\tef & log g & [Fe/H] & Comp.  &  Reference   \\
     &       &        & Star   &              \\
\hline
 \noalign{\smallskip}
5857 & 4.47  & +0.23   & Sun     & This Paper \\     
5764 & 4.40  & +0.21   & Sun     & Jenkins et al. (2008) \\
5767 & 4.45  & +0.16   & Sun     & Soubiran  et al. (2008) \\
5857 & 4.47  & +0.22   & Sun     & Valenti \& Fisher (2005) \\
5673 & 4.22  & -0.01   & Sun     & Pasquini et al. (1994)  \\
5793 & 4.50  & +0.20   & Sun     & Boesgaard \& Friel (1990)  \\
5793 & 4.60  & +0.24   & Sun     & Abia  et al. (1988)     \\
5793 &  --   & +0.16   & Sun     & Boesgaard \& Budge (1988)  \\
5860 & 4.40  & +0.28   & Sun     & Rebolo et al. (1986)    \\
5793 & 4.50  & +0.19   & Sun     & Cayrel de Strobel et al. (1985)    \\
5860 & 4.40  & -0.09   & Sun     & Cayrel de Strobel  et al. (1981) \\

\hline
\hline
\noalign{\smallskip}
\end{tabular}
\end{table}

HD10700 is a high proper-motion dwarf star, studied 
by many authors. Simbad lists 756 references as of February 2012, ranging from 
chromospheric activity 
studies (Jenkins et al. 2006, Mart\'{i}nez-Arn\'{a}iz et al. 2010) through to 
stellar oscillation studies 
(Teixeira et al. 2009).  In Table \ref{_thd} we show the iron 
abundances studied by different authors for HD10700.

\begin{table}
\centering
\caption{\label{_thd} Temperatures, gravities and abundances used in studies 
of HD10700.}
\begin{tabular}{ccccl}
\hline
\noalign{\smallskip}
  \tef & log g & [Fe/H] &Comp. & Reference   \\
       &        &         & Star &             \\
\hline
  5383  &  4.59  &  -0.55    &  Sun  &  This Paper   \\
 5377  &  4.53  &  -0.49     &  Sun  &    Mashonkina et al. (2011) \\ 
   5522  &  4.50  &  -0.37    &  Sun  &  Jenkins et al. (2008) \\
   5344  &  4.45  &  -0.52    &  Sun  &  Soubiran et al. (2008) \\ 
   5310  &  4.44  &  -0.52    &     -     &  Sousa et al. (2008) \\  
   5264  &  4.36  &  -0.50    &  Sun  &  Cenarro et al. (2007) \\  
   5283  &  4.59  &  -0.52   &   Sun  &  Valenti \& Fischer (2005) \\
   5320  &  4.30  &  -0.50    &  Sun  &  Castro et al. (1999) \\  
   5330  &  4.30  &  -0.59    &  Sun  &  Tomkin et al. (1999) \\
   5500  &  4.32  &  -0.38    &  Sun  &  Mallik et al. (1998) \\ 
   5250  &  4.65  &  -0.46    &  Sun  &  Arribas et al. (1989) \\ 
   5143  &  3.60  &  -0.60    &  Sun  &  Barbuy et al. (1989) \\   
   5305  &  4.32  &  -0.66    &  Sun  &  Gratton (1989) \\  
   5250  &  4.50  &  -0.58    &  Sun  &  Abia et al. (1988) \\  
   4990  &  4.50  &  -0.56    &  Sun  &  Francois (1986) \\  
   5305  &  4.33  &  -0.49    &  Sun  &  Steenbock (1983) \\  
   5362  &  4.59  &  -0.34    &  Sun  &  Hearnshaw (1974) \\  
   5538  &    -   &  -0.13    &   Sun  &  Herbig (1965) \\  
   5196  &    -   &  -0.39    &   Sun  &  Pagel et al. (1964) \\  
   5305  &    -   &  -0.39    &   Sun  &  Pagel (1963) \\  

\hline
\hline
\noalign{\smallskip}
\end{tabular}
\end{table}

The overall goal of this paper is to outline our new procedure to measure 
precise and accurate 
iron abundances, microturbulent velocities and rotational velocities for 
solar-type stars through analysis of high 
resolution stellar spectra of the Sun, HD1835 and HD10700.  

\section{Procedure}

\subsection{Observed data}

  \begin{table*} 
    \centering 
     \caption{\label{_list} Parameters of stars of our interest} 
      \begin{tabular}{c|c|c|c|c|c} 
      \noalign{\smallskip} 
       \hline 
        \noalign{\smallskip} 
	 
	Star & \tef (VF05) & log g (VF05) & [Fe/H] (VF05) & $V_r$ (km/s) & \vsini (km/s) \\ 
	   \noalign{\smallskip} 
	       \hline 
        \noalign{\smallskip}
the Sun &5777                &  4.44  &  0  &                          &    1.7  \\
HD1835  &  5857.0$\pm$22   & 4.47 & +0.22  & -2.7$\pm$1.8 (Jenkins et al. 2011)   &  6 (VF05), 8 (Jenkins et al. 2008) \\
HD10700 &  5283$\pm$22     & 4.59 & -0.52 &    -16.9$\pm$1.6  (Jenkins et al. 2011)    &  2 (VF05)   \\
	 
	\hline       
	\hline 
	\noalign{\smallskip} 
	\end{tabular} 
	\end{table*} 

The adopted parameters for our stars are listed in Table \ref{_list}.
The spectra and all calibration data were observed using
the Fibre-fed Extended Range Optical Spectrograph (FEROS)
mounted on the MPG/ESO - 2.2m telescope on the La Silla
site in Chile. The exposure times were long enough to ensure that the 
spectra were observed with S/N ratios of well over 
150 in the continuum around the iron lines at
6100 \AA\ at the operating resolution of FEROS ($R\sim$48$'$000).  All 
calibration files needed for the reduction of the stellar spectra
(flat-fields, bias and arc frames) were obtained at the beginning
and end of each nights observing, following the standard ESO
calibration plan. The reduction of all the spectra followed the 
standard reduction techniques described by Jenkins et al. (2008). 

The observed spectrum of the Sun (Kurucz et al. 1984) is used as the  
reference star. The resolution of the solar spectrum is much higher,
($R=$ 100$'$000), however profiles of observed lines 
in the solar spectrum are broadened by macroturbulence.
We adopt a gaussian for the macroturbulence profile 
with a FWHM = 0.1\AA\ at 6000\AA. This corresponds to 
V$_{\rm macro}$ = 2.2~\kms which agrees well with the reference value
(Gurtovenko \& Sheminova 1986).

\subsection{Model atmospheres}

HD1835 and HD10700 are stars which are similar to the Sun, validating 
the Sun as our reference star. For each star we computed 
plane-parallel model atmospheres in LTE, with no energy divergence, using 
the SAM12 program (Pavlenko 2003), which is a modification of 
ATLAS12 (Kurucz 1999). In this study  as ``the zeroth approach''
we adopt values given by Valenti \& Fischer (2005) 
for both stars HD1835 and HD10700, i.e.. \tef/log g/[Fe/H] = 5857/4.47/+0.22 and
5383/4.75/-0.36, respectively, for faster convergence of the code. For the Sun we used a model atmosphere of
5777/4.44/0.00. All models were recomputed with the new abundances 
found by ourselves.
Chemical equilibrium is computed for molecular species assuming LTE and we use 
the opacity sampling approach from Sneden (1976) to account for absorption of 
atoms, ions and molecules (see more details in Pavlenko 2003). The 1-D 
convective mixing length theory, modified by Kurucz (1999) in ATLAS12, was 
used to account for convection. The computed model atmospheres are available on 
the 
web\footnote{ ftp://ftp.mao.kiev.ua/pub/users/yp/MA2010}. 

\subsection{Synthetic spectra}

Synthetic spectra are calculated with the WITA6 program (Pavlenko 1997), 
using the same approximations and opacities as SAM12. 
To compute the synthetic spectra we use line lists taken from VALD2 
(Kupka et al. 1999). The shape of each atomic line is determined 
using a Voigt function and all damping constants are taken from line 
databases, or computed using Unsold's approach (Unsold 1954). 
A wavelength step of $\Delta\lambda$ = 0.025 \AA\ is employed in the 
synthetic spectra computations to match our observed 
FEROS spectrum.   It is worth noting that different atomic species provide 
different contributions to the formation of the spectrum of 
solar-like stars. In particular, iron lines dominate the solar spectrum. 

\subsection{\label{_AvS} Best fit parameters selection}

In our work we used two spectroscopic datasets to carry out the 
abundance analysis. The first one
consists of the spectroscopic data taken from the VALD (Kupka et al. 1999). 
For any absorption line we identify the atom, molecule or ion, its 
central wavelength, 
$\lambda_{\rm o}$, the oscillator strength of the transition $gf$, the 
excitation potential  
of the lower level \EE of the corresponding transition and the 
damping constants $C_{\rm 2}, C_{\rm 4}, C_{\rm 6}$.
The second one contains the list of pre-selected spectral regions ($N_{L}$) governed 
by absorption of the atom and/or ion of interest.

\subsubsection{\label{_vvt}Pre-selection of spectral features in the solar spectrum}

We used the Sun as a template star to verify our procedure. Indeed, the solar 
abundances were determined by many authors using different procedures and the 
results do not differ  very little.     

The first step of our selection procedure was to compute a spectrum of 
the Sun taking into account only lines of elements of interest. 
A comparison with the 
observed spectrum provides a list of spectral features in which absorption
of our element 
dominates. 
Blending 
by other lines was accounted for directly where the contribution of other 
lines can be
estimated in a simple way from the analysis of the ratio $r(Fe)/r(Sun)$, where
$r(Fe)$ and $r(Sun)$ are residual fluxes from the continuum normalisation
 in our theoretical spectrum of iron lines 
and the observed spectrum of the Sun. Then, from the analysis of the shape of 
the $r(Fe)/r(Sun)$ ratio 
we determine the central wavelengths of every feature and spectral region in 
which we compare the theoretical and observed spectra to determine the iron 
abundance. The spectral range of comparison was chosen to be approximately 
between -0.1\AA\ and +0.1\AA\ from the 
blue and red edge of the corresponding spectral feature, a method shown to be 
effective in Jenkins et al. (2008).  We do note that the synthetic spectra were 
computed across a broader spectral region of approximately $\pm$5 \AA.

From a comparison of the observed and computed spectra we obtained a list of strong 
features governed by the absorption of a given element.  
Strong features were selected to reduce the possible effects of noise. 
As a limit for our strong lines we choose the value $r(Fe)$ = 0.8.

\subsubsection{\label{_fe}Fits to pre-selected spectral features}  

For each preselected spectral region that contains an absorption line we are 
interested in, we carry out the fit to the observed spectrum.  The synthetic 
spectra that are computed across the selected spectral region are convolved 
with profiles that match the instrumental broadening and that take into account 
rotational broadening.  For instrumental broadening we adopt a Gaussian profile and 
the rotational broadening was treated following the scheme by Gray (1976).  The 
instrumental broadening takes care of the spectrograph resolution, and \vsini is a 
parameter used to get the best fit to the observed profiles of any feature in our spectrum.

For every feature we find the minimisation parameter:
$$ S_{\rm l} = \sum (1-r_{\nu}^s/r_{\nu}^{\rm o})^2/N_o$$
here $N_o$ is the number of points in the observed spectrum across the 
pre-selected spectral region,
 $r_{\nu}^{\rm o}$ and $r_{\nu}^{\rm s}$ are residual fluxes in the 
observed and computed spectra, respectively.
 Formally a minimum $S_l$ determines our solution, i.e.. the abundance of an element 
governed by the associated absorbing
feature on the adopted set of abundances $X_i= log (N(X_i)), i=1,..., N_a$,
where $N_a$ is the dimensionality of the abundance grid. 
 
Our measurements are affected by some errors of a differing nature:
uncertainties in $gf$, blending, presence of noise in the observed spectrum,
etc. Every fitted spectral feature provides one abundance  
$X_l$ on $min ~ S_l$, $l$ = 1, $N_L$ 
we compute mean abundance 

$$X_o = (\sum X_l)/L)$$
and a formal standard error $\sigma_o$

$$\sigma_o=\sqrt{\sum(X_l -X_o)^2/(N_l*(N_l-1))} $$ 
where $N_l$  is the total number of the fitted spectral lines. 

It is worth noting
that sometimes not all pre-selected features provide a minimum $S$ on our grid of abundances,
therefore $N_l \leq N_L$. In other words, if the minimum of $S$ cannot be found 
in the adopted abundance range, or any 
absorption feature is too weak in the observed spectrum, the line
was excluded from the following consideration.

Additionally, we compute the abundance of elements averaged over 1$\sigma$,
i.e. we averaged all abundances across \\ $X_o + 1\sigma_o$: 

$$X_s =\sum(X_l/\sigma_l)/\sum(1/\sigma_l)$$

In this paper we adopt $N_a$ = 30 with a step of 0.05, 
$N_v$ = 6 with a step of 0.5 km/s and
$N_r$ = 11 with a step of 0.4 - 0.5 km/s.

In some details our procedure is similar to that used by 
Jones et al. (2002) and Pavlenko \&
Jones (2002), but in this case we individually fit the synthetic spectra to 
the pre-selected features in the observed spectrum.

Our procedure allows us to  obtain
the iron abundances from the fits to Fe I and Fe II independently. 
The Fe I/Fe II = log N(Fe I)-log N(Fe II),  where N(Fe I) and N(Fe II) are
respectively the iron abundance determined from the fits to Fe I and Fe II
lines, is used to verify the 
log g of the model atmospheres used in the fits. 

 It is worth noting a few important points:

-- since we deal with blending by absorption lines of other elements, 
which allows us to include strong blended features in the observed spectrum. 

-- rotational velocity (\vsini) is determined for every point (log N(Fe), \Vt)
of our synthetic spectral grid for every spectral line.

\subsection{\label{_alg} General algorithm of solution}

\subsubsection{Dependence of abundance of iron vs. excitation potential of 
absorption lines}

In our analysis
we use a set of lines of different excitation potentials of the lower level of the 
corresponding radiative transition \EE. In the first approach their response on 
temperature in the line-forming region of the stellar atmosphere can be described
by $exp(-E''/kT)$, where the labels have the conventional meaning. Lines of different 
\EE show different response to T (and respectively, on \tef). In our case we work with 
blends in the stellar spectra too. They are treated explicitly. 
 If a few lines 
of the given element form the feature then we use the 
``effective'' excitation  potential 
$$E^{''} = -ln (\Sigma (g_i*exp(-E_i/kT))/\Sigma(g_i))*kT$$ 
here values and labels have 
the conventional meaning, $i = 1,..., N$. Here $N$ is the total number of lines of
the given element that formed the feature.

Ideally, for the properly determined \tef we should not obtain any 
dependence of
log N(Fe) on \EE for absorption lines of both Fe I and Fe II. In other words 
the absence of dependence of log N(Fe) = f(\EE) we 
consider as a sufficient condition for the determination of 
microturbulent velocity and abundance. 
 It it worth noting that a similar 
approach was used by other authors in the past 
(see fig.8 in Mashonkina et al 2011).

\begin{figure}
   \centering
   \includegraphics[width=88mm]{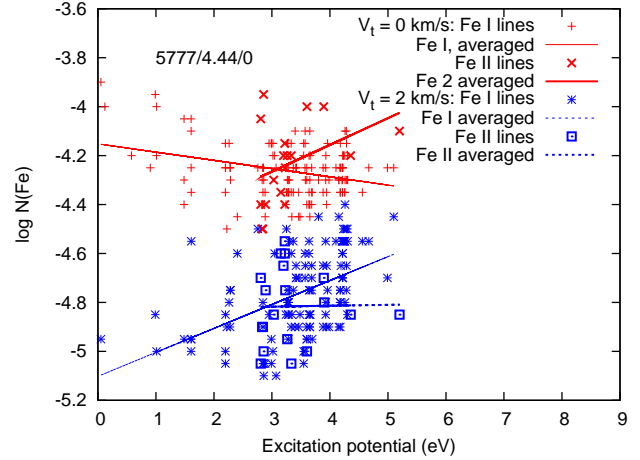}
    \caption{\label{_59ss} Dependence of the iron abundance determined from the 
    fits to the observed
     Fe I and Fe II features in the observed spectrum of the Sun vs. 
     excitation potential computed for a model atmosphere with 
parameters 5777/4.44/0.}
   \end{figure}

In Fig. \ref{_59ss} we plot the dependence of Fe I and Fe II abundances in the spectrum of the Sun versus their excitation 
potentials. These computations were carried out for a model atmosphere of 5777/4.44
and microturbulent velocities of \Vt = 0 km/s and 2 km/s. The results of the approximation of the dependence of 
log N(Fe) = f (\EE) is highlighted by the straight line fits to the data in the figure.

Even visual inspection of Fig. \ref{_59ss} provides a good estimation
of microturbulence velocity in the solar atmosphere 0 $<$ \Vt $<$ 2 km/s.
A more complete 
set of the approximations obtained for \Vt =0, 0.5, 1, 1.5, 2, 2.5 km/s will be shown 
in section \ref{_RES} for our star of interest.

\subsubsection{The self-consistent solution}

We  pay special attention to the ionisation equilibrium
Fe I/Fe II. Lines of these ions are abundant in the spectra of solar-like stars.
Furthermore, the spectroscopic data for Fe I
and Fe II lines are more accurate and complete in comparison with other ions.

To determine the realistic parameters of our stars we developed and used 
the following custom algorithm:

1) Using the fits to the profiles of Fe I and Fe II lines separately, we compute 
dependences of log N(Fe) = f(\Vt) for the model atmosphere 
with adopted \tef on a fixed grid of microturbulence velocities.
In general log N(Fe I) = log N(Fe II) is a pre-requisite to find the proper 
\tef.

2) In some cases we obtain a few solutions log N(Fe I) = log N(Fe II) for 
different \tef and \Vt, and \vsini. To select the best solution we test if there 
is any dependence of derived abundances on 
excitation potential of lower level of the corresponding spectral features,
as described in section \ref{_fe}.
 
3) After the determination of log N(Fe), \tef, \Vt 
we determined the abundances of other elements.

4) If we obtain large differences in abundances in comparison with the
results of the previous iteration then we recompute the model atmosphere and repeat
all these procedures.

It is worth noting that in comparison with the former papers we 
reduced the number of free parameters.
Namely, we determined log N(Fe), \Vt and \vsini from the best fit to the 
observed spectrum. Then, we used an averaging procedure across the 1$\sigma$ 
error parameter space to obtain a statistically significant result. Finally, 
the best solution 
was found from the best agreement of all three parameters on the grid 
of parameters. Furthermore, the metallicity of the model atmosphere agrees 
with the 
fitting results, i.e.. we found our final solution in a few successive 
approximations taking 
into account the variability of the abundances and other parameters  
in only a few iterations. 

\section{\label{_RES}Results}

\subsection{Formal test of the procedure: abundances, \Vt, and \vsini
for the Sun}

To test our procedure we employed a few model atmospheres with parameters 
5777/4.44, 5777/4.50, 5527/4.50 and 6027/4.50.  We then 
determined log (Fe) for  the adopted grid of microturbulent velocities
\Vt = 0, 0.5, 1, 1.5, 2, 2.5 km/s 
using the fits to Fe I and Fe II lines in the observed solar spectrum. 
Error bars 
were computed following the standard procedure described in 
section \ref{_fe}.

\begin{figure*}
   \centering
   \includegraphics[width=85mm]{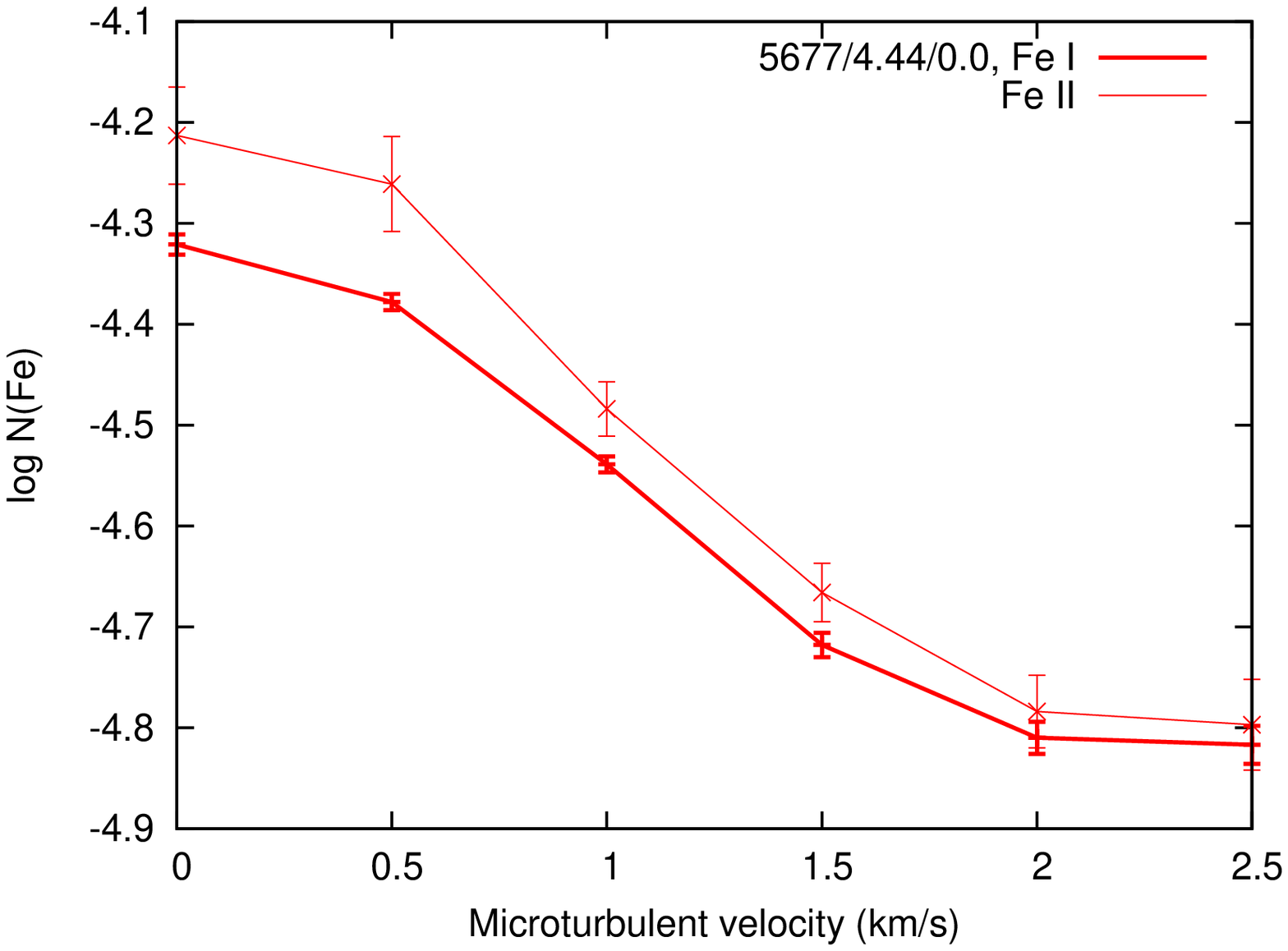}
   \includegraphics[width=85mm]{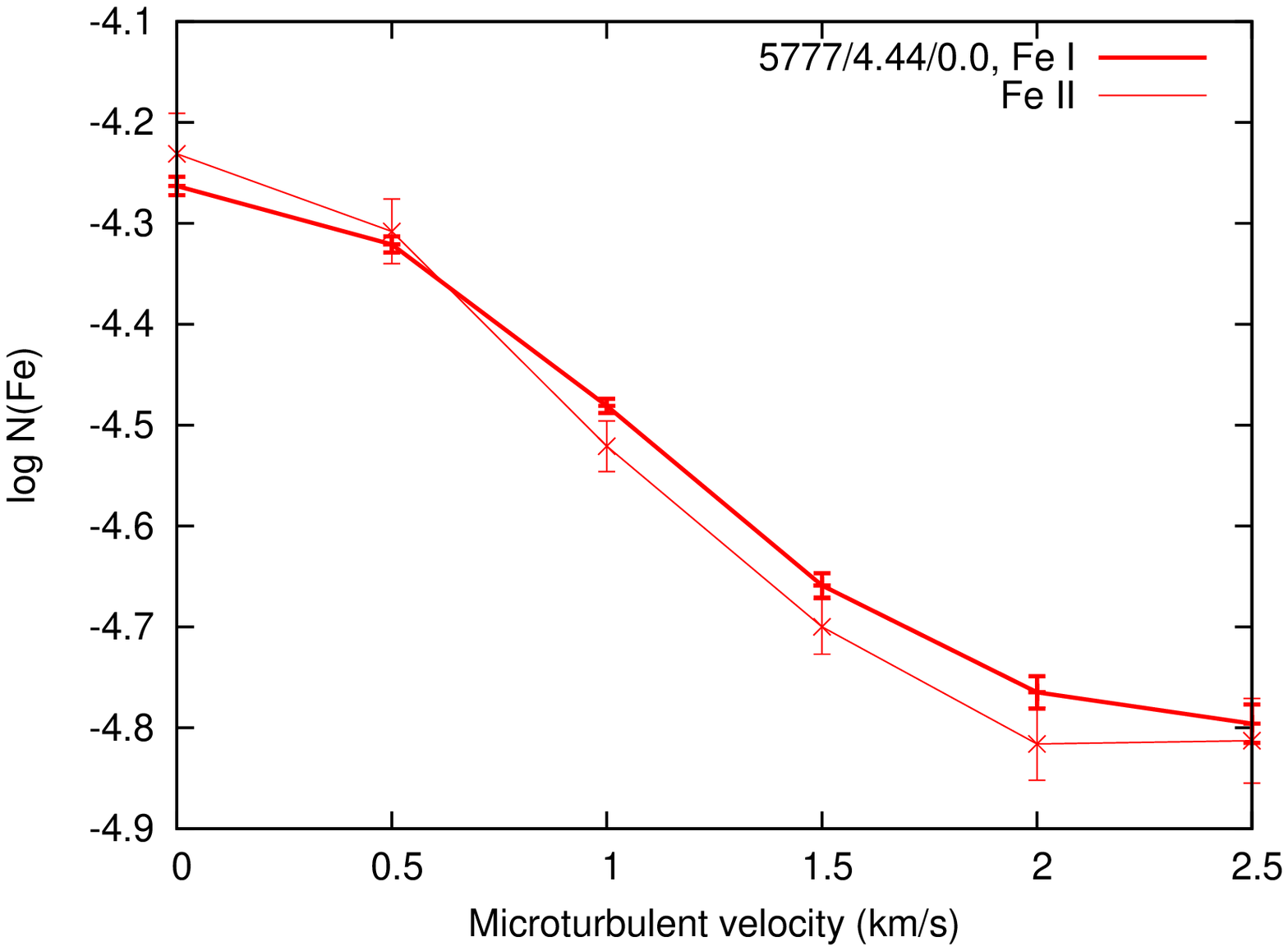}   
\includegraphics[width=85mm]{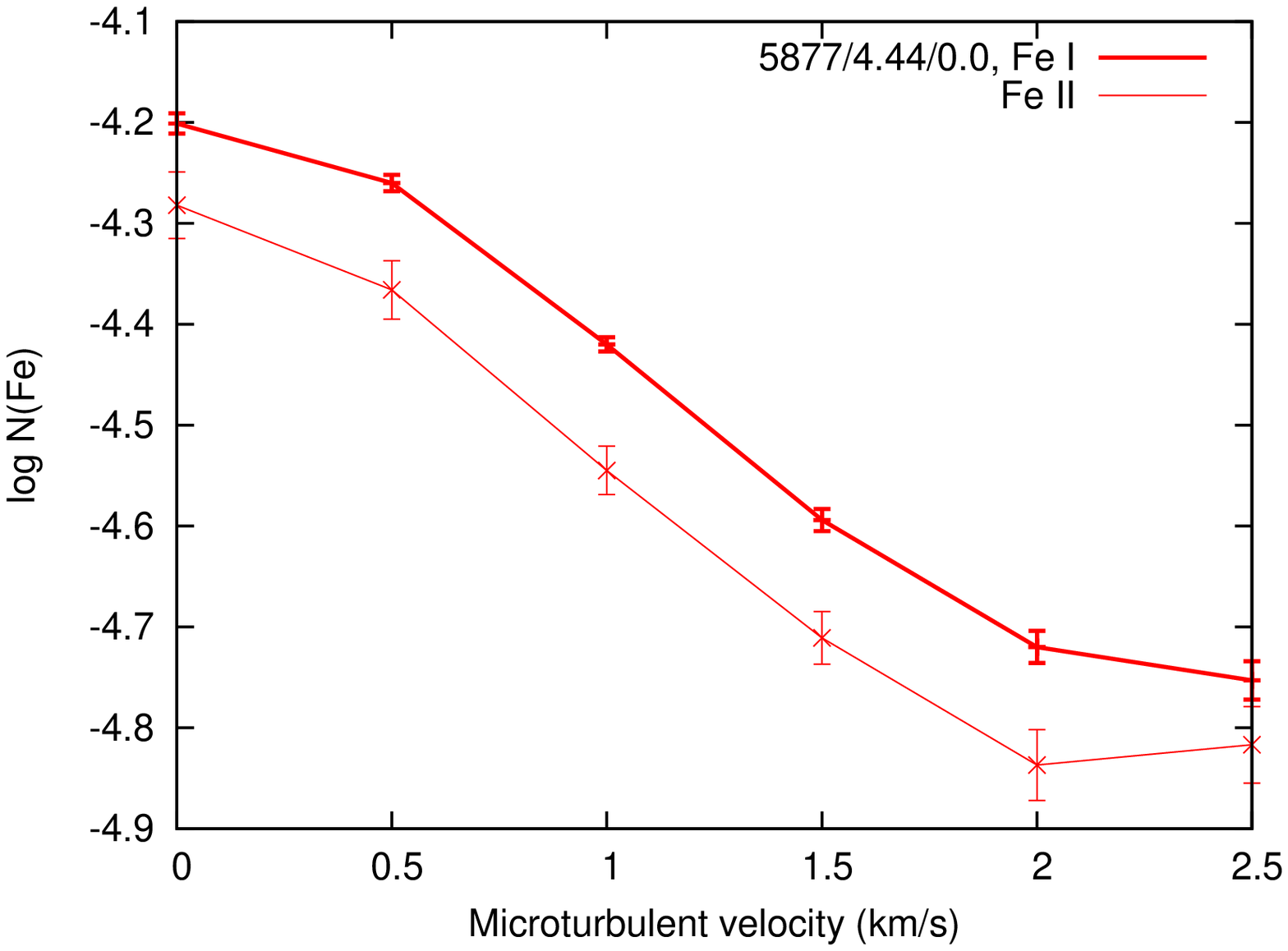}
\includegraphics[width=85mm]{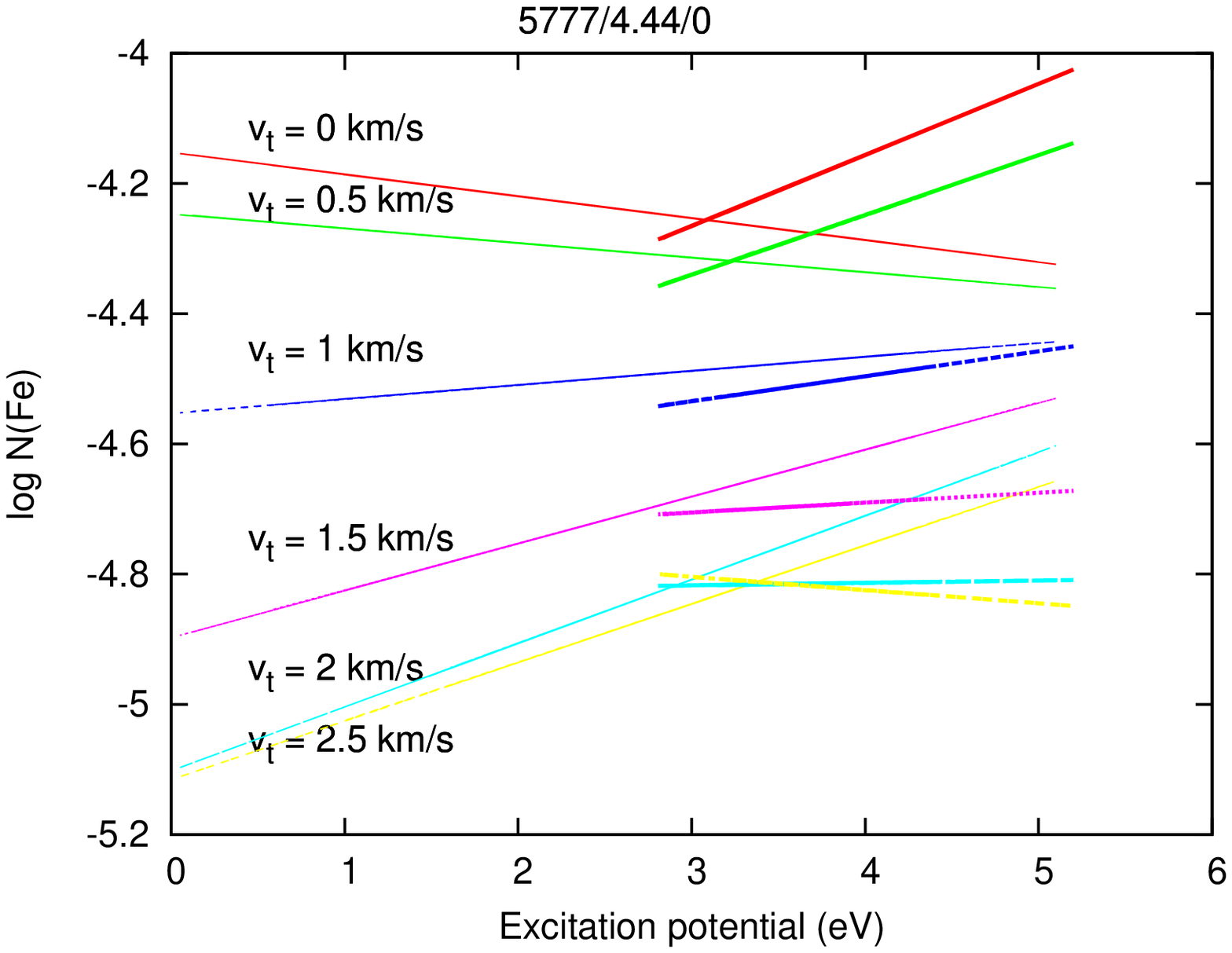}   
    \caption{\label{_fes} Abundances of iron determined from the fits to the 
    observed
     Fe I and Fe II features in the observed spectrum of the Sun.
     Model atmospheres are 5677/4.44, 5777/4.44, and 5877/4.44 for the panels 
at the top left, top right, and lower left, respectively.  The lower right 
panel shows the 
dependence of log N(Fe) vs. \EE of Fe I and Fe II lines shown by thin and 
thick lines, respectively, 
used to computed  the corresponding abundances for the model 
atmosphere of 5777/4.44.}
   \end{figure*}

Results  of the iron determinations for three model atmospheres are shown in 
Fig. \ref{_fes}. As was expected, only model atmospheres with \tef = 5777 K 
provide similar values of iron abundances obtained from the fits of Fe I and Fe II
lines.  Model atmospheres of lower \tef provide overestimated abundances of Fe II ions
in the atmosphere of the Sun and underestimate the abundances of Fe I, respectively, and 
vice versa, model atmospheres of higher \tef, underestimate abundances of Fe II and 
overestimate Fe I.

\begin{table*}
\centering
\caption{\label{_ts} Abundances in the atmosphere of the Sun as a star}
\begin{tabular}{ccccccc}

\noalign{\smallskip}
\hline

       &        \multicolumn{3}{c} {\Vt = 0.75 km/s}               &    \multicolumn{3}{c}{\Vt = 1.25 km/s}               \\

\hline
\noalign{\smallskip}

      & Mean $X_o$ & Averaged $X_s$ & \vsini & $Mean X_o$ & Averaged $X_s$ &\vsini \\
\hline
\noalign{\smallskip}

 Ca I  &-5.592 $\pm$  0.020 & -5.609 $\pm$  0.114 &  1.300 $\pm$  0.392 & -5.767 $\pm$  0.019 & -5.777 $\pm$  0.105 &  1.133 $\pm$  0.342\\
 Cr I  &-6.307 $\pm$  0.021 & -6.440 $\pm$  0.143 &  1.758 $\pm$  0.265 & -6.473 $\pm$  0.018 & -6.391 $\pm$  0.136 &  1.740 $\pm$  0.262\\
 Fe I  &-4.395 $\pm$  0.007 & -4.394 $\pm$  0.005 &  1.644 $\pm$  0.144 & -4.566 $\pm$  0.009 & -4.514 $\pm$  0.004 &  1.598 $\pm$  0.140\\
 Fe II &-4.418 $\pm$  0.023 & -4.498 $\pm$  0.112 &  1.805 $\pm$  0.426 & -4.603 $\pm$  0.026 & -4.744 $\pm$  0.166 &  1.616 $\pm$  0.381\\
 Ni I  &-5.745 $\pm$  0.013 & -5.621 $\pm$  0.054 &  1.743 $\pm$  0.247 & -5.880 $\pm$  0.016 & -5.787 $\pm$  0.043 &  1.555 $\pm$  0.220\\
 Ti I  &-7.027 $\pm$  0.019 & -7.022 $\pm$  0.078 &  1.652 $\pm$  0.312 & -7.237 $\pm$  0.020 & -7.365 $\pm$  0.119 &  1.541 $\pm$  0.291\\
 Ti II &-6.847 $\pm$  0.037 & -6.602 $\pm$  0.176 &  1.944 $\pm$  0.381 & -7.097 $\pm$  0.038 & -6.863 $\pm$  0.136 &  1.841 $\pm$  0.361\\

\noalign{\smallskip}
\hline\hline

\noalign{\smallskip}
\end{tabular}
\centering
\end{table*}

It is worth briefly noting two points:

\begin{itemize}

\item Visual inspection of the plots similar to that shown 
in Fig \ref{_fes} can be used to determine the proper \tef.

\item  We see good agreement in iron abundances obtained from the fits of Fe I 
and Fe II lines across the wide range of microturbulent velocities for the model 
atmospheres 5777/4.44 and 5777/4.5. 

\end{itemize}

To solve the last problem we investigated the obtained abundances 
versus excitation potential of the main absorbing lines. 
We approximate the obtained dependence of computed
abundances versus \EE  by a linear function using least-squares fitting.
The dependence log N(Fe) vs.. \EE computed  for a model atmosphere of the Sun of 
5777/4.44/0.0 and different microturbulent velocities 
 is shown in the lower panel of Fig. \ref{_fes}. Results of fits to Fe I lines
 show the absence of the log N(Fe) vs. \EE for \Vt = 0.75 km/s and log N(Fe) =
 -4.40. It is worth noting that 
Sheminova \& Gadun (2010) 
analysed the Sun at high resolution (R=200000) 
and found \Vt =0.8 
$\pm$ 0.2 km/s.  Nevertheless, our value of iron abundance corresponds well 
enough with 
Grevesse and Anders (1979) log N(Fe) = -4.37 and Gurtovenko and Kostik (1989)
log N(Fe)= -4.40. However, our fits to Fe II lines provide a little lower 
abundance log N(Fe)= -4.6.
The differences of iron obtained from the fits to Fe I and Fe II lines are 
discussed in many papers (see the most recent paper Mashonkina et al. 2011 and 
discussion therein). On the other hand, our sample of Fe I lines is 
larger, therefore 
the results obtained from the fits to the the neutral ion lines in the 
observed spectrum are more robust.

Using the derived estimations of \tef, log g, and \Vt we obtained 
abundances of other elements using the aforementioned scheme.
The comparison of our results with known abundances for the Sun is presented
 in Table \ref{_tfin}.
For most abundances where we found satisfactory agreement, the residuals do not exceed 
0.1 dex. However, we found a difference of 0.16 dex against the Cr I 
abundance measured by Grevese and Anders (1989). On the other hand 
our value agrees well with the Gurtovenko \& Kostik value log N(Cr) = -8.07.
Therefore, we estimate our accuracy of abundance determination as $<$ 0.1 dex.
This level of precision provides us with a good opportunity to investigate absolute 
differences in the abundances of stars like the Sun. 

To investigate the dependence of abundances on the adopted microturbulent velocity 
we repeat our abundance determination procedure for the case of \Vt = 1.25 km/s
(see Table \ref{_ts}). For all elements we obtained similar results, i.e.
difference in the adopted microturbulent velocity of 0.5 km/s provide changes in the 
obtained abundances by a factor of 0.2 dex.

In Table \ref{_ts} we show the rotational velocity of the Sun determined from the 
fits of our features to the observed spectrum of the Sun as a star (Kurucz et al. 1984).
For the solar spectrum we adopt an effective resolution R = 70$'$000, the formal resolution
 limited by the presence 
of macroturbulent  velocity $V_{ma}$ = 1 - 2.6 km/s (Gurtovenko \& Sheminova 1986,
Sheminova \& Gadun 2010).
Our determined \vsini = 1.6 $\pm$ 0.3 agrees well with the 
known rotational velocity of the Sun
\vsini $\sim$ 1.85 $\pm$ 0.1  km/s (see Bruning 1984), and corresponds well with 
\vsini =1.63 km/s obtained by Valenti \& Fisher (2005). 

One may expect that
the determined rotational velocity for the Sun depends on the adopted 
resolution. Indeed, even a slightly lower resolution of R = 60$'$000 will 
overestimate the contribution from macroturbulence
into the total broadening and underestimate the \vsini = 1.2 $\pm$ 0.3 km/s.

However, it does not affect the results of abundance determinations 
because the observed spectra are broadened by the combined profile formed by 
rotation + macroturbulence + instrumental broadening, where the broadening 
shape is dominated by the instrumental profile.
  
\subsection{HD1835}

\begin{table*}
\centering
\caption{\label{_ta} Abundances in the atmosphere of HD1835}
\begin{tabular}{ccccccc}
\noalign{\smallskip}
\noalign{\smallskip}
\hline
       &        \multicolumn{3}{c} {5807/4.47/+0.2, \Vt = 0.75 km/s}       &    \multicolumn{3}{c}{5857/4.47/+0.2, \Vt = 0.75 km/s}           \\  
\noalign{\smallskip}
\noalign{\smallskip}
\hline
      & Mean log [X/H] & log [X/H] averaged & \vsini  & Mean log [X/H] & log [X/H] averaged & \vsini \\

\hline
\noalign{\smallskip}
\noalign{\smallskip}
  Ca I & -5.262 $\pm$ 0.036 & -5.322 $\pm$ 0.213 & 7.000 $\pm$ 2.111& -5.229 $\pm$ 0.040 & -5.282 $\pm$ 0.121 &7.042 $\pm$2.123    \\
  Cr I & -6.016 $\pm$ 0.034 & -6.017 $\pm$ 0.121 & 7.560 $\pm$ 1.181& -6.000 $\pm$ 0.032 & -6.203 $\pm$ 0.083 &7.500 $\pm$1.217    \\
  Fe I & -4.135 $\pm$ 0.020 & -4.110 $\pm$ 0.008 & 7.137 $\pm$ 0.626& -4.124 $\pm$ 0.018 & -4.111 $\pm$ 0.006 &7.094 $\pm$0.632    \\
  Fe II& -4.117 $\pm$ 0.060 & -4.146 $\pm$ 0.060 & 6.722 $\pm$ 1.630& -4.121 $\pm$ 0.057 & -4.222 $\pm$ 0.116 &6.737 $\pm$1.588    \\
  Ni I & -5.482 $\pm$ 0.036 & -5.469 $\pm$ 0.033 & 7.596 $\pm$ 1.064& -5.445 $\pm$ 0.034 & -5.434 $\pm$ 0.036 &7.539 $\pm$1.066    \\
  Ti I & -6.815 $\pm$ 0.058 & -6.719 $\pm$ 0.066 & 7.352 $\pm$ 1.442& -6.728 $\pm$ 0.059 & -6.614 $\pm$ 0.084 &7.429 $\pm$1.430    \\
  Ti II& -6.663 $\pm$ 0.069 & -7.037 $\pm$ 0.152 & 7.407 $\pm$ 1.453& -6.643 $\pm$ 0.070 & -7.040 $\pm$ 0.141 &7.444 $\pm$1.460    \\

\noalign{\smallskip}                                                         
\hline                                                                       
\hline                                                                       
\end{tabular}

\centering
\end{table*}

\begin{figure*}
   \centering

      \includegraphics[width=85mm]{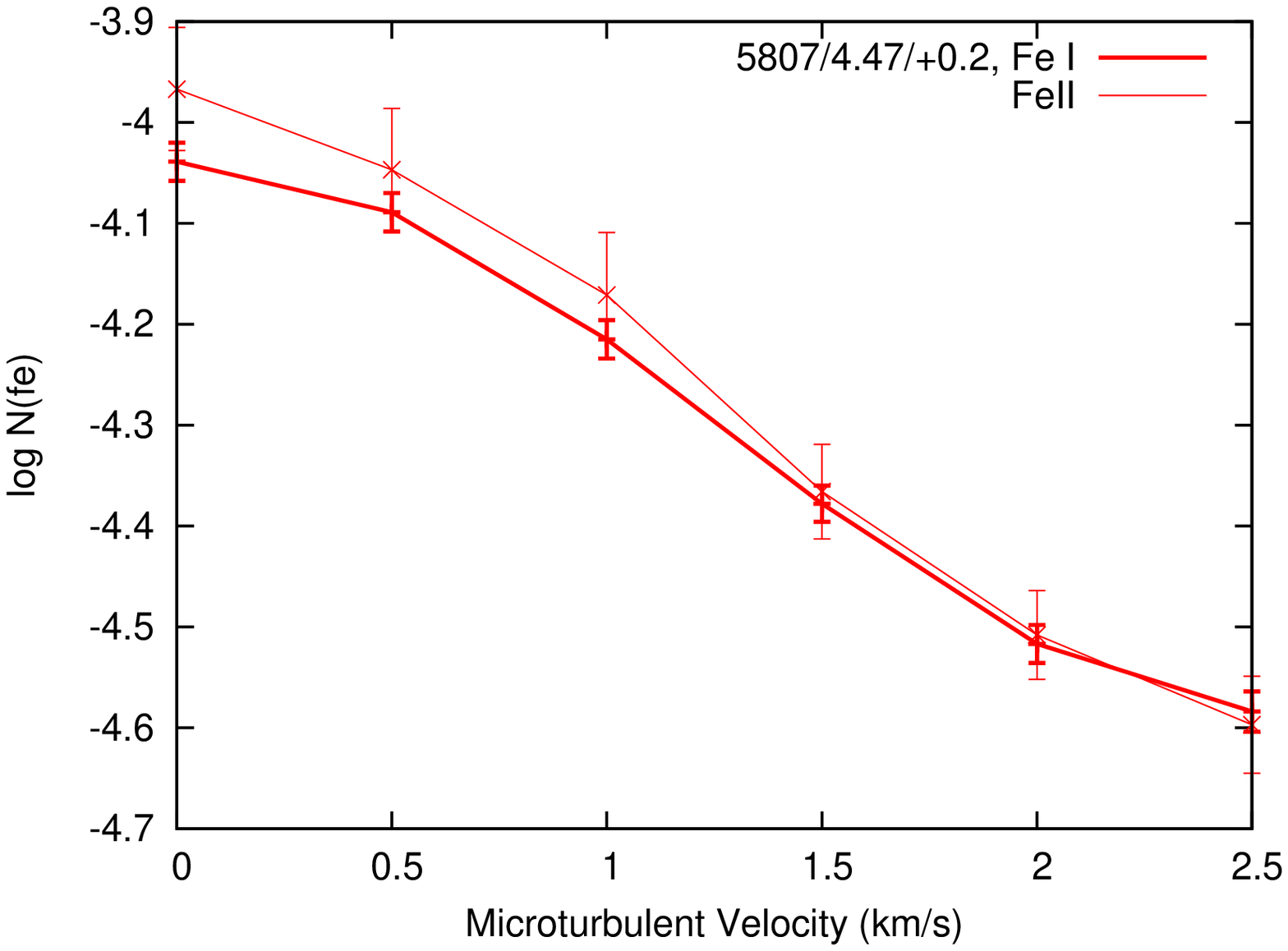}
      \includegraphics[width=85mm]{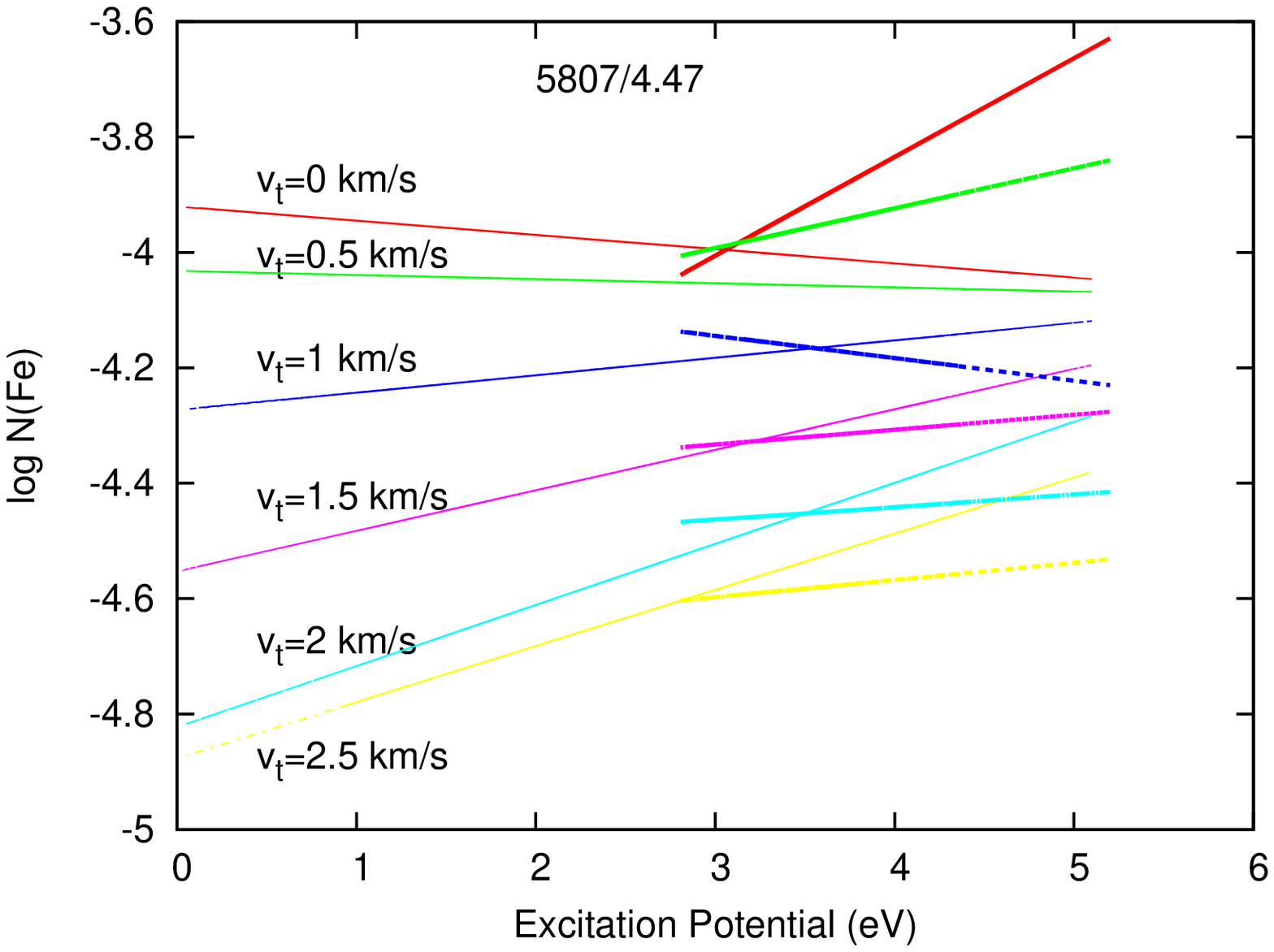}
\caption{\label{_fea} Left: abundances of iron determined from the fits 
on the synthetic stectra computed for the  5807/4.47/+0.2 model atmosphere
to the observed
     Fe I and Fe II features in the observed spectrum of HD1835.
Right: the dependence log N(Fe) vs. \EE of Fe I and Fe II lines shown 
by thin and 
thick lines, respectively. }
   \end{figure*}
   
For HD1835 we obtained good agreement between abundances obtained from 
fitting the Fe I and Fe II lines over a wide range of \Vt for model atmospheres of \tef 5807
and 5857 K. However, the
additional proof of the dependence of the obtained abundances on \Vt, 
shown at the bottom right panel of Fig. \ref{_fea}, gives 
better results of log N(Fe) = -4.18 for the 5807/4.47/+0.2 model atmosphere and 
\Vt = 0.75 km/s.  The abundances of other elements determined with these 
parameters are given in the Table \ref{_ta}.

For HD1835 we investigated the dependence of our abundance determination on 
\tef. Differences of 100K in the adopted effective 
temperature changes the determined abundances by only 0.1 dex. As for 
the Sun, the formal accuracy of our \tef determination for HD1835  is of 
the order $\sim\pm$50K. 

Using lines of different elements and ions
we obtained the rotation velocity \vsini = 7.2 $\pm$ 0.5 km/s. This value 
agrees with the determination from previous authors 
(see Jenkins et al. 2008 and
references therein).

\section{HD10700}

\begin{figure*}
   \centering

      \includegraphics[width=85mm]{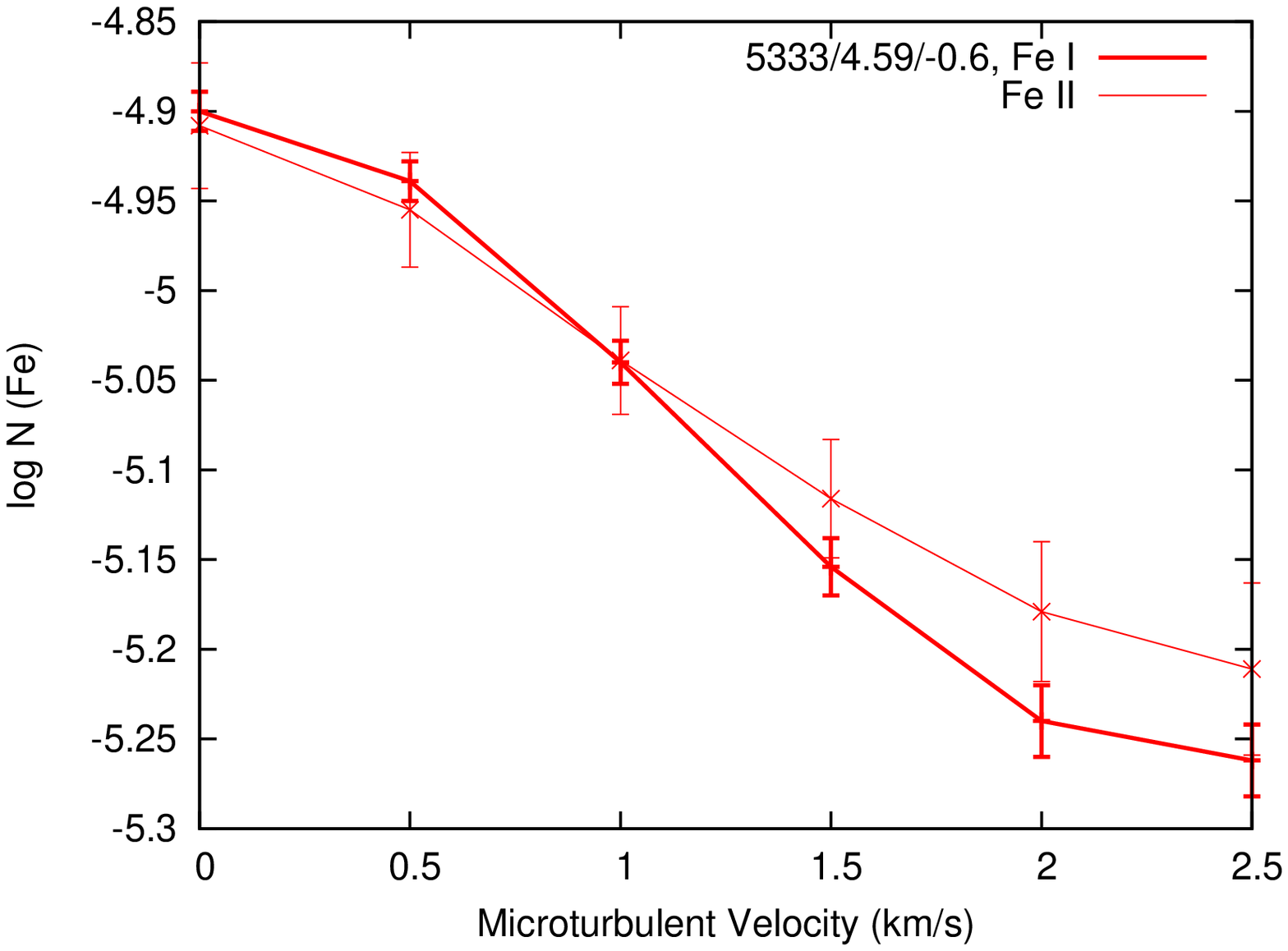}
      \includegraphics[width=85mm]{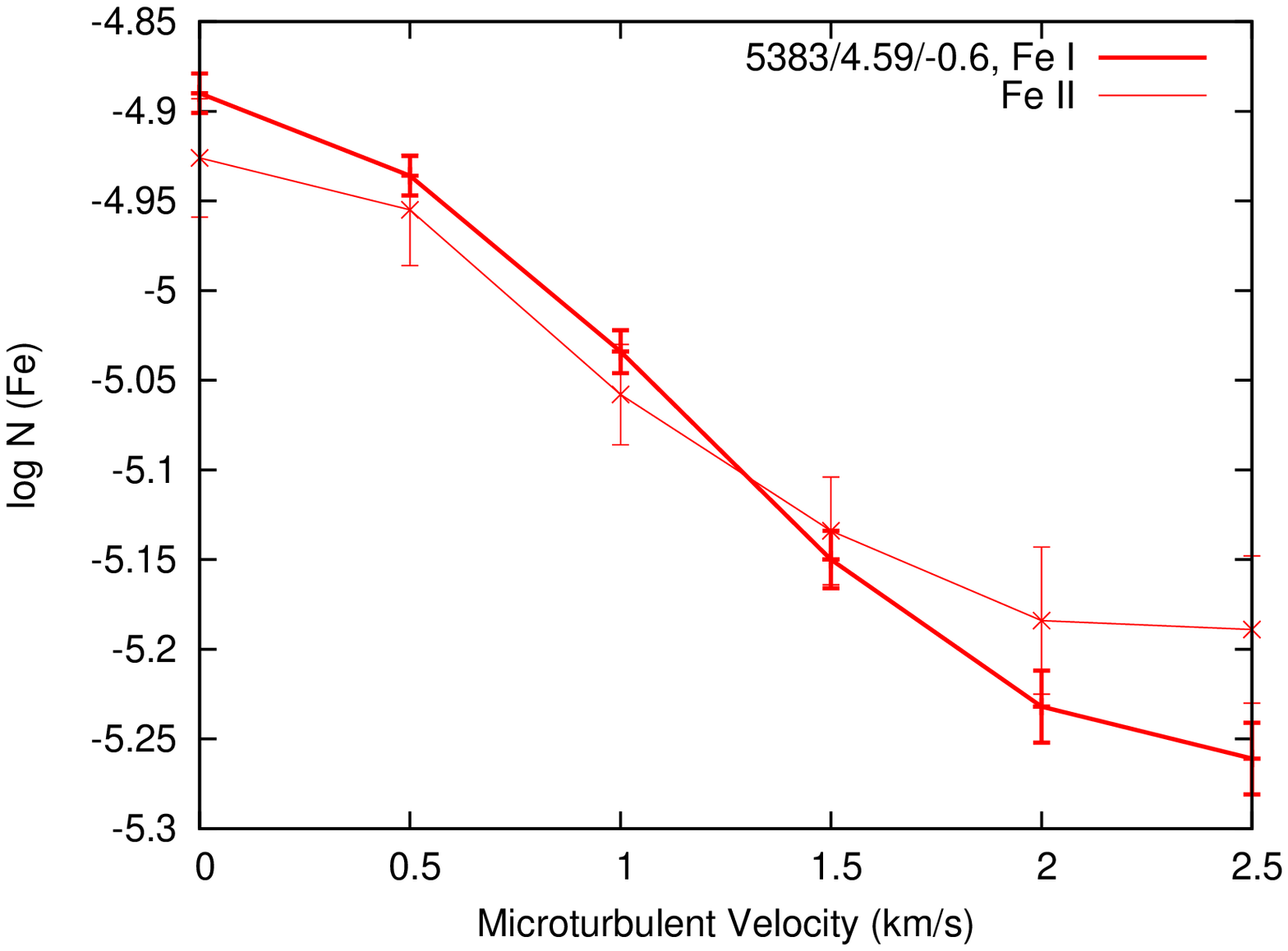}
       \includegraphics[width=85mm]{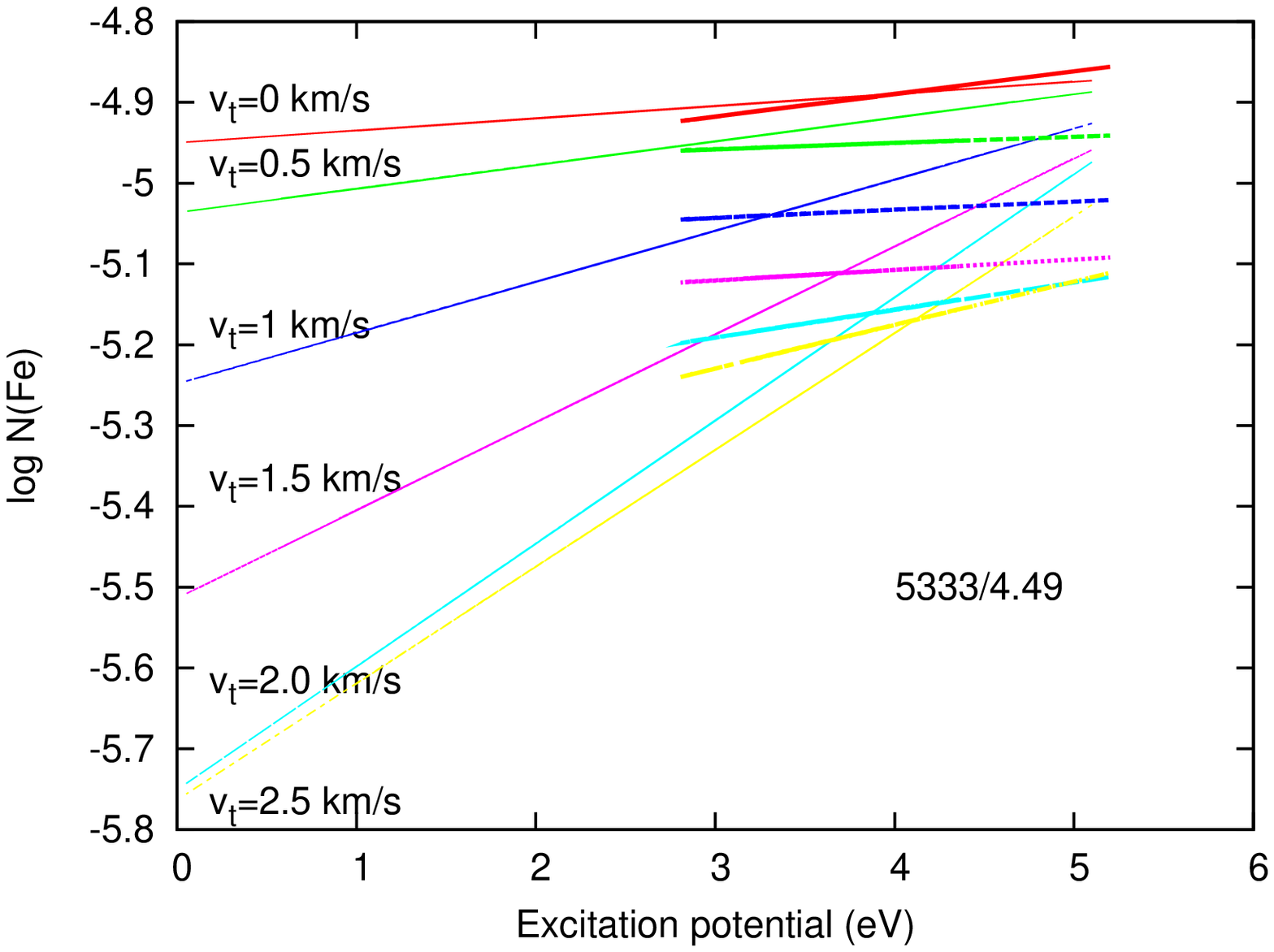}
      \includegraphics[width=85mm]{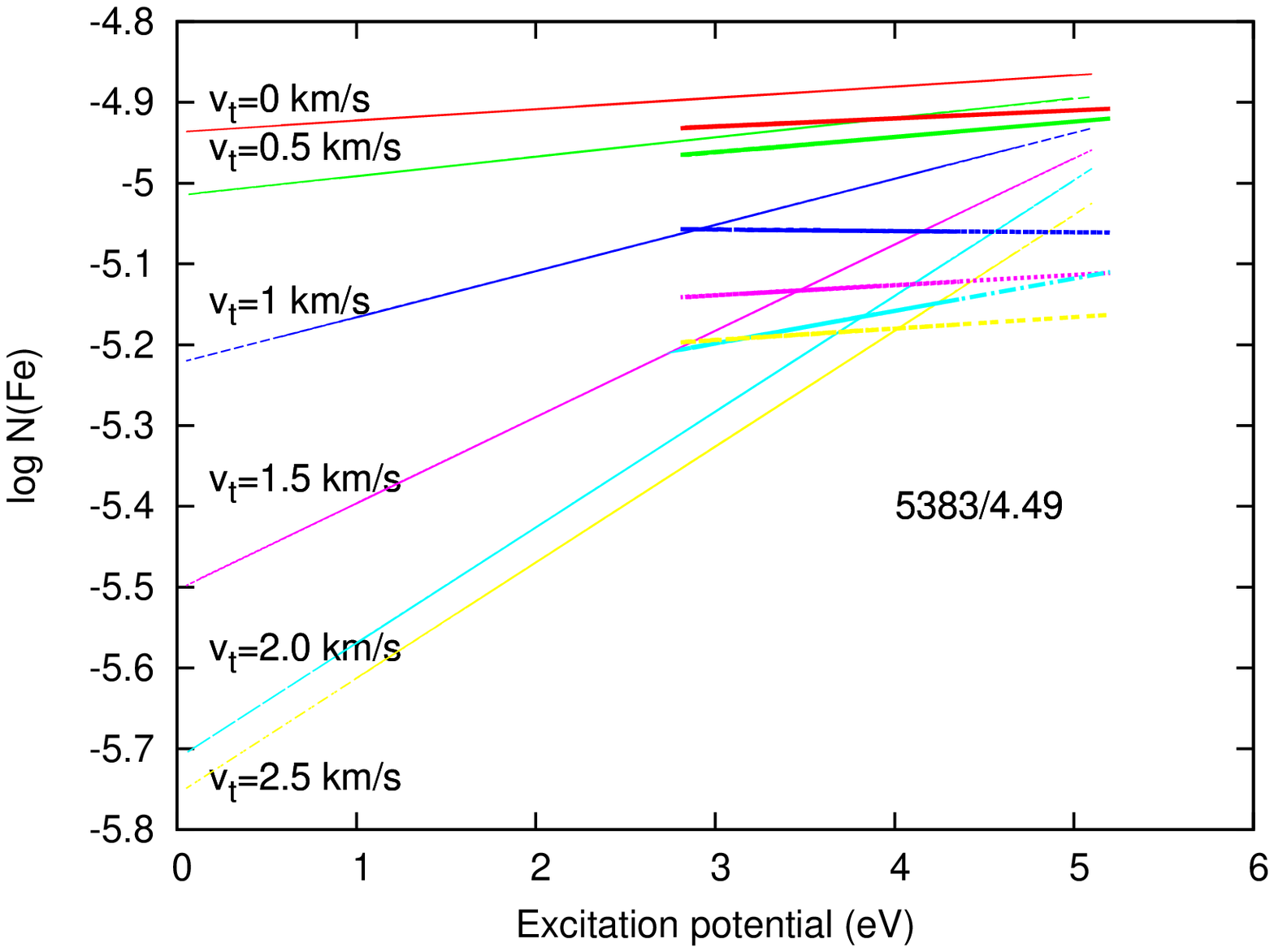}
\caption{\label{_feb} Top: abundances of iron determined from the fits 
of synthetic spectra computed for the  5333/4.49/-0.5 and 5383/4.90/-0.5 
model atmospheres  to the observed
     Fe I and Fe II features in the observed spectrum of HD10700.
Bottom: the dependence log N(Fe) vs. \EE of Fe I and Fe II lines is
shown by thin and thick lines, respectively.}
   \end{figure*}

For the metal deficient star HD10700 we computed a small grid of model
atmospheres of \tef = 5133 - 5500 K with a step of 50K and a log g = 4.30 - 4.90 with 
a step of 0.3.  All models have a metallicity of [Fe/H] = -0.6. 
First of all, following Valenti 
\& Fisher (2005) we adopted log g =4.59 and obtained a good enough solution 
for two model atmospheres of \tef = 5333 and 5385 K. Both Fe I and Fe II 
solutions provide log N(Fe) = -4.95 $\pm$ 0.05 for a rather low \Vt = 0.5 
km/s, see (Fig. \ref{_feb}).  A small dependence of log N(Fe) = f (\EE) is
obtained for the Fe I lines, which cannot be removed by changes of \tef and log g.
Again, the rotational velocity obtained from the fits of lines of different
elements is in the range 2 - 2.7 km/s.

We then proceeded to investigate the dependence of our results on 
log g. In general, the solution based on the Fe~I/Fe~II ionisation equilibrium shows a
systematic increase of \tef in the case of model atmospheres of higher log g,
and vice versa, a decrease of log g for the lower \tef.

\begin{figure*}
   \centering

      \includegraphics[width=85mm]{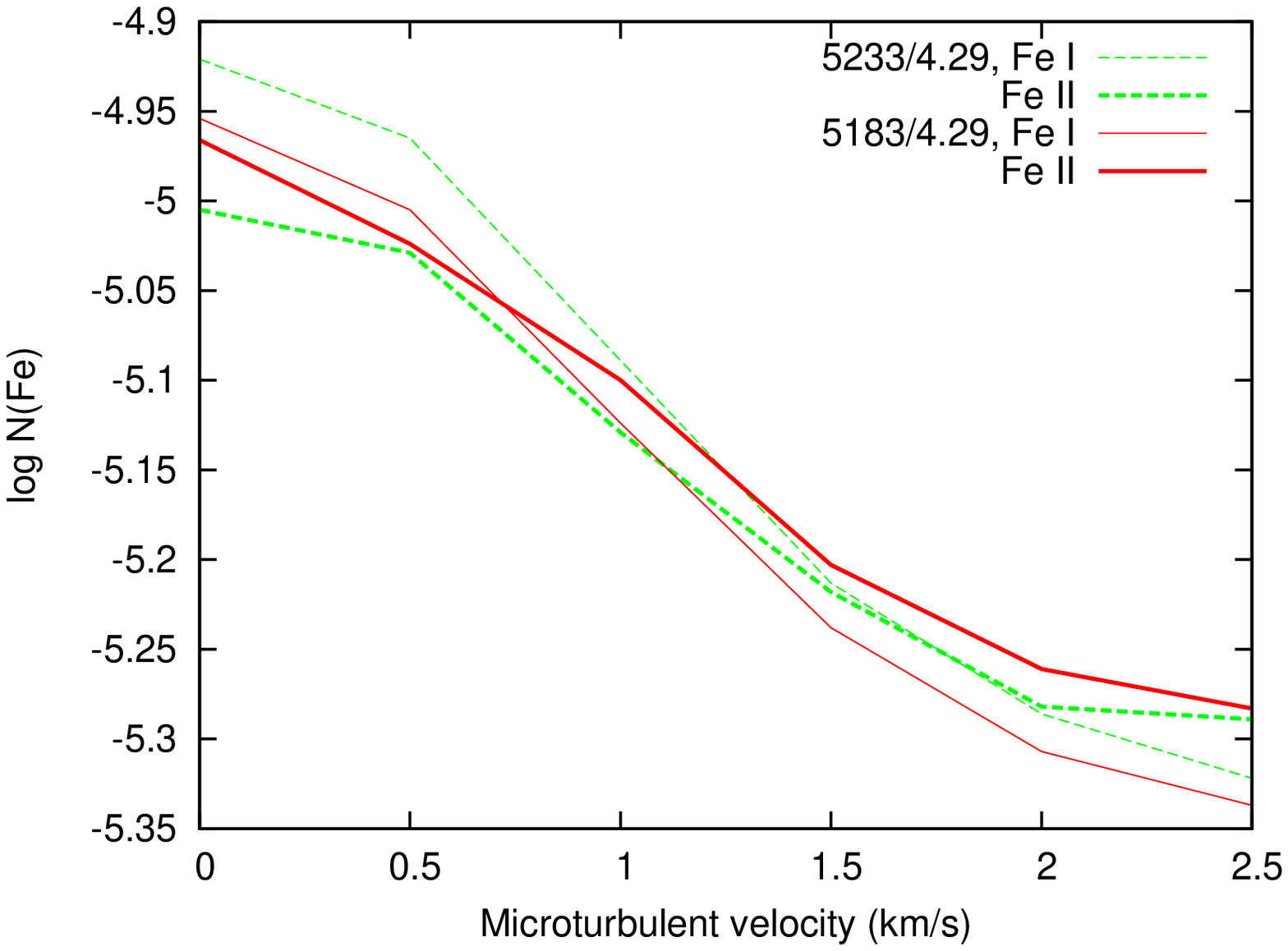}
      \includegraphics[width=85mm]{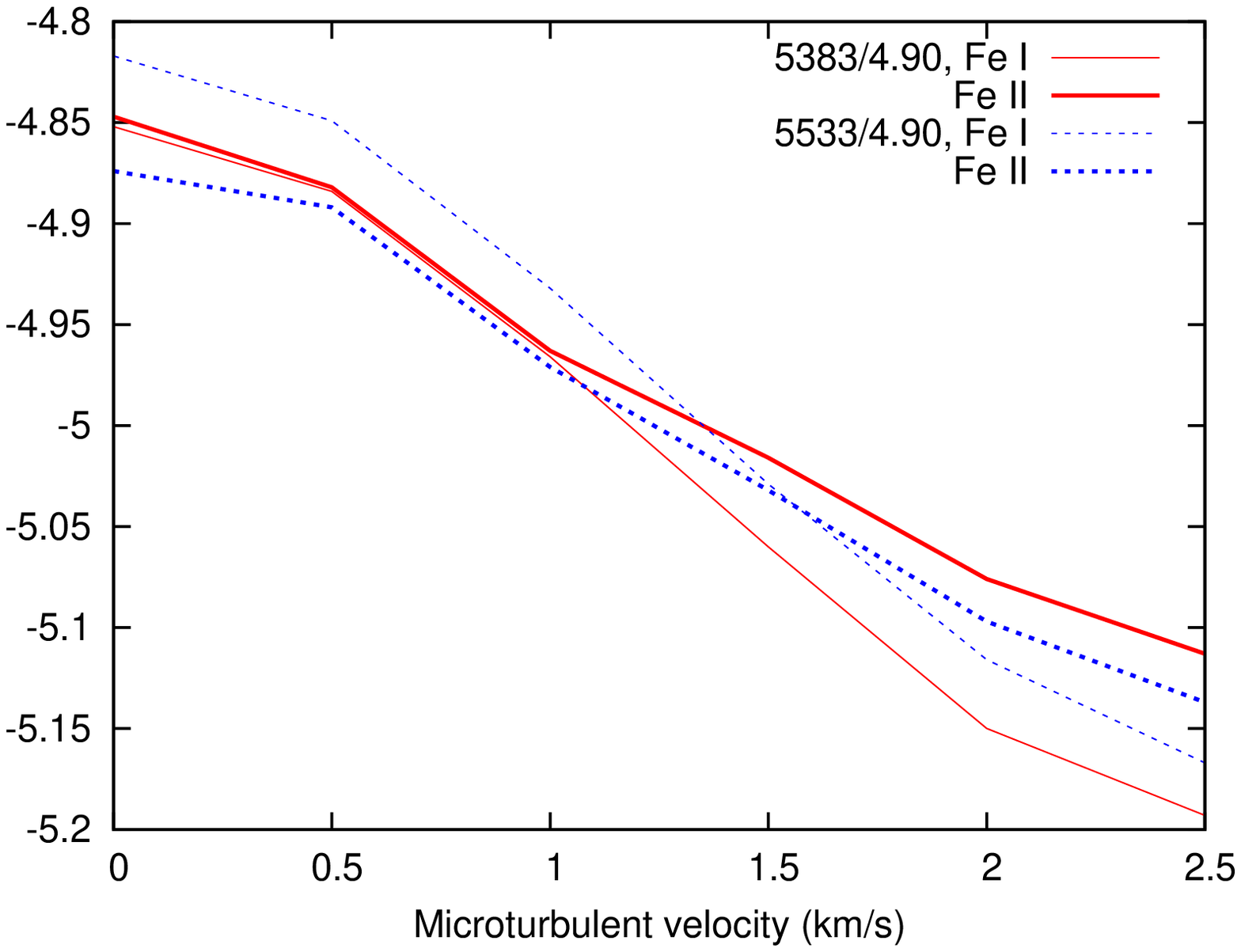}
       \includegraphics[width=85mm]{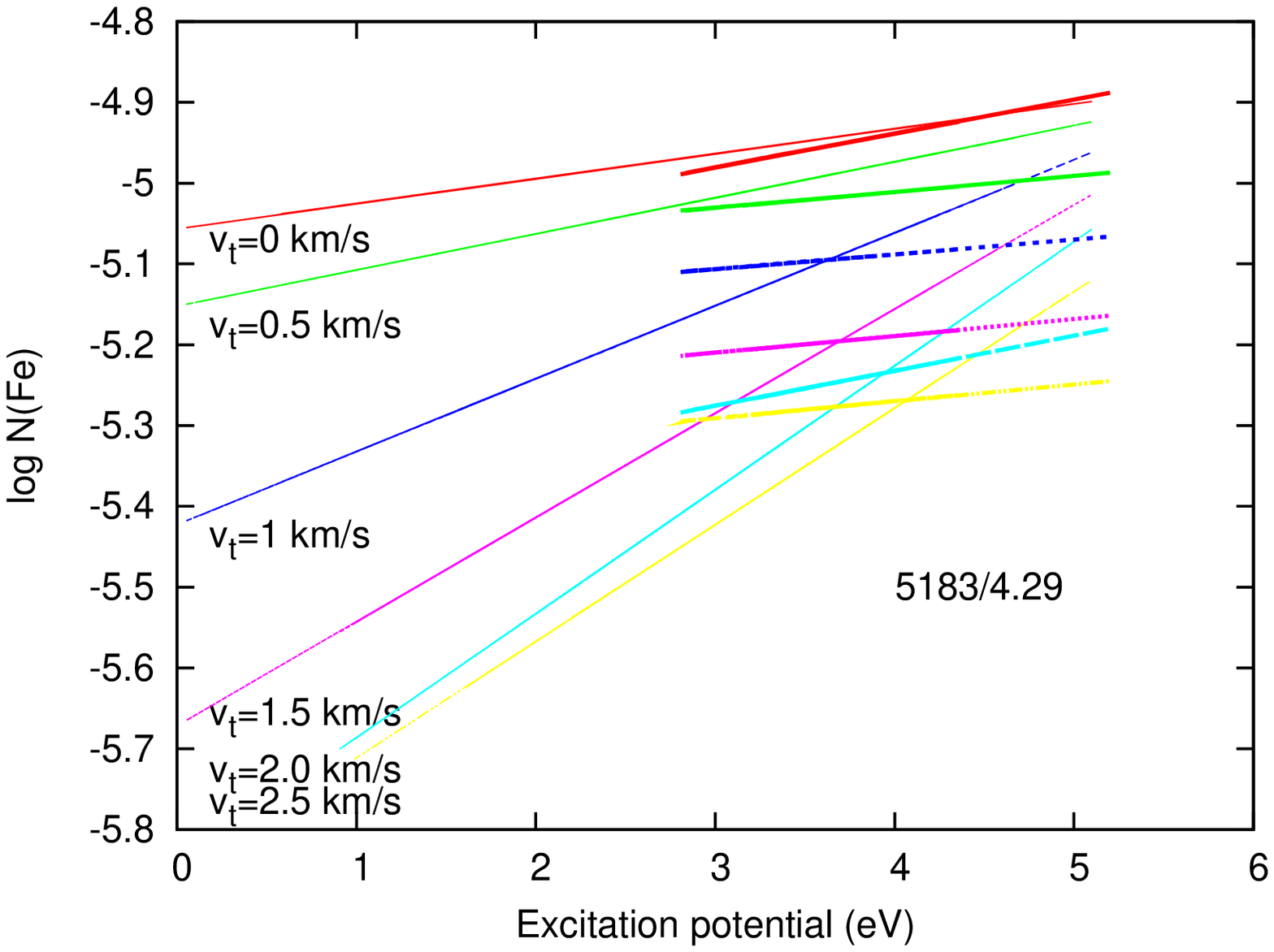}
      \includegraphics[width=85mm]{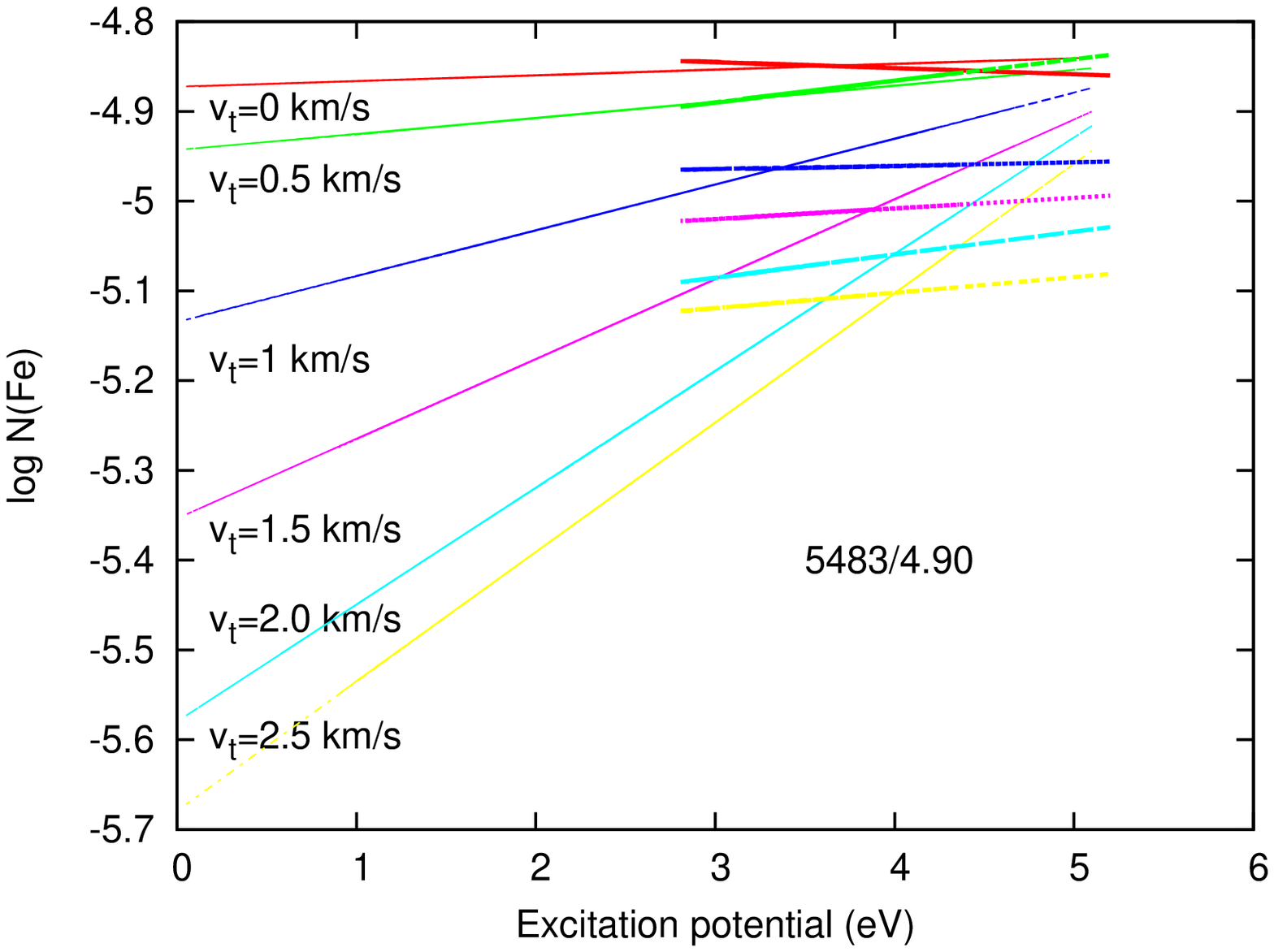}
\caption{\label{_logg} Top: 
abundances of iron determined from the fits
of synthetic spectra computed for model atmospheres of
different \tef  and log g to the observed
     Fe I and Fe II features in the observed spectrum of HD10700.
       Bottom: the dependence log N(Fe) vs. \EE of Fe I and Fe II lines is 
shown for the 5183/4.29 and 5483/4.90 
model atmospheres by thin and thick lines, respectively.}
   \end{figure*}

Nevertheless, we are able to obtain a good agreement in the Fe I/Fe II for 
cooler model atmospheres of lower log g and hotter models of larger \tef
(see Fig. \ref{_logg}). Still using the log N(Fe) vs.. \EE proof we found that
the hotter model atmospheres of higher gravities provide more reasonable 
results. Again, we obtain a self-consistent solution for the case of the rather
low \Vt, and, respectively, higher metallicity.

\begin{table*}
\centering
\caption{\label{_tb}Abundances in the atmosphere of HD10700}
\begin{tabular}{ccccccc}
\noalign{\smallskip}
\noalign{\smallskip}
\hline
       &        \multicolumn{3}{c} {5333/4.59/-0.6, \Vt=0.5 km/s}               &    \multicolumn{3}{c}{5383/4.49/-0.6, \Vt=0.5 km/s}           \\  
\noalign{\smallskip}
\noalign{\smallskip}
\hline
      & Mean log [X/H] & log [X/H] averaged & \vsini  & Mean log [X/H] & log [X/H] averaged & \vsini\\
\hline
\noalign{\smallskip}
\noalign{\smallskip}

    Ca I  & -5.942 $\pm$ 0.034 &-5.958 $\pm$ 0.131  &  2.250 $\pm$  0.678& -5.933 $\pm$  0.032 & -5.943 $\pm$  0.108 &  2.167 $\pm$  0.653 \\
    Cr I  & -6.837 $\pm$ 0.024 &-6.877 $\pm$ 0.041  &  2.733 $\pm$  0.412& -6.814 $\pm$  0.024 & -6.846 $\pm$  0.056 &  2.700 $\pm$  0.407 \\
    Fe I  & -4.939 $\pm$ 0.011 &-4.911 $\pm$ 0.005  &  2.622 $\pm$  0.230& -4.936 $\pm$  0.011 & -4.910 $\pm$  0.005 &  2.580 $\pm$  0.226 \\
    Fe II & -4.955 $\pm$ 0.032 &-4.910 $\pm$ 0.080  &  2.632 $\pm$  0.620& -4.955 $\pm$  0.031 & -4.905 $\pm$  0.097 &  2.711 $\pm$  0.639 \\
    Ni I  & -6.254 $\pm$ 0.013 &-6.257 $\pm$ 0.061  &  2.000 $\pm$  0.283& -6.253 $\pm$  0.013 & -6.250 $\pm$  0.051 &  2.000 $\pm$  0.283 \\
    Ti I  & -7.298 $\pm$ 0.035 &-7.412 $\pm$ 0.107  &  2.643 $\pm$  0.509& -7.271 $\pm$  0.034 & -7.492 $\pm$  0.183 &  2.571 $\pm$  0.495 \\
    Ti II & -7.290 $\pm$ 0.038 &-7.239 $\pm$ 0.105  &  2.500 $\pm$  0.521& -7.281 $\pm$  0.038 & -7.257 $\pm$  0.104 &  2.521 $\pm$  0.526 \\

\noalign{\smallskip}
\hline\hline
\end{tabular}
\centering
\end{table*}

\begin{table*}
\centering
\caption{\label{_tfin}Abundances in atmospheres of our stars  relative to the 
conventional solar abundances of GA89, i.e. Grevesse \& Anders (1989) and GK89, i.e. Gurtovenko \& Kostik (1989).}
\begin{tabular}{ccccccc}

\noalign{\smallskip}
\noalign{\smallskip}
\hline
      &  \multicolumn{2}{c} {5777/4.44/0.0, \Vt=0.5 km/s} &  \multicolumn{2}{c} {5807/4.47/+0.2, \Vt=0.75 km/s}
 &  \multicolumn{2}{c} {5333/4.59/-0.6, \Vt=0.5 km/s} \\

\noalign{\smallskip}
\noalign{\smallskip}
\hline
          &   GK89 &  GA89      & GK89 &  GA89      &  GK89              &  GA89 \\
\noalign{\smallskip}
\noalign{\smallskip}
\hline

Ca I  &  0.068  &  0.088 $\pm$   0.020 &  0.398 &   0.418 $\pm$   0.036 &  -0.273 &  -0.253 $\pm$   0.032  \\
Cr I  &  0.133  &  0.063 $\pm$   0.021 &  0.424 &   0.354 $\pm$   0.034 &  -0.374 &  -0.444 $\pm$   0.024  \\
Fe I  &  0.005  & -0.025 $\pm$   0.007 &  0.265 &   0.235 $\pm$   0.020 &  -0.536 &  -0.566 $\pm$   0.011  \\
Fe II & -0.018  & -0.048 $\pm$   0.023 &  0.283 &   0.253 $\pm$   0.060 &  -0.555 &  -0.585 $\pm$   0.031   \\
Ni    &  0.075  &  0.045 $\pm$   0.013 &  0.338 &   0.308 $\pm$   0.036 &  -0.433 &  -0.463 $\pm$   0.013   \\
TiI   & -0.047  &  0.023 $\pm$   0.019 &  0.165 &   0.235 $\pm$   0.058 &  -0.291 &  -0.221 $\pm$   0.034    \\
TiII  &  0.133  &  0.203 $\pm$   0.037 &  0.317 &   0.387 $\pm$   0.069 &  -0.301 &  -0.231 $\pm$   0.038    \\

  \noalign{\smallskip}
  \hline\hline
  \end{tabular}
  \centering
  \end{table*}

\subsubsection{Final remarks}

The big advantage of our procedure is the possibility for independent 
proof of the measured results. First of all, our 
abundances and \vsini show a rather weak dependence on the adopted
parameters of the model atmosphere.
We show that changes in the abundances are small (0.1 dex) if one
is close to the appropriate temperature ($\pm$ 100K) and gravity ($<$ 0.3 dex) 
and thus our strategy
is valid, at least for the solar-like stars of
the solar neighborhood. Formal errors
of our determination in most cases do not exceed 0.1 dex, a value that is 
good enough to determine abundances in the atmospheres of a large 
number of stars and perform statistical analyses on such samples.

\section{Discussion}

We have developed a numerical scheme which allows the determination of atomic 
abundances, 
microturbulent velocities and rotational velocities in stellar atmospheres
in the framework of a self-consistent approach. Minimisation of the differences 
in 
the profiles of computed and observed lines allow us to carry out the 
determination of
abundances of elements for neutral and ionic species separately. 
This allows us to provide an independent verification of the adopted effective 
temperatures, gravities and metallicities for the star in question. 

We used the Sun as a template star to select the most appropriate 
lines of atoms and ions. Blending effects are accounted for directly, i.e.. 
synthetic spectra are computed with the inclusion of all lines. This is important 
because even in the case of the solar spectrum we cannot fit to all lines without 
considering blends.
To reduce possible numerical errors we account for 
only a part of each spectral feature which shows a strong enough 
dependence on the 
adopted abundance. It is worth noting that the procedure of line selection
from the solar spectrum was done within a narrow range of abundances. However, 
even in this case the fraction of blended lines can affect our results.

To derive abundances in the atmospheres of HD1835 and HD10700 we used the 
same list of spectral features governed by absorption lines
 determined from fits to the solar spectrum. Our procedure works well enough 
 for the case of metal rich stars (HD1835) and metal deficient stars (HD10700),
 for slow rotators (the Sun, HD10700), and intermediate fast rotators (HD1835).
 It is worth noting that for these 3 stars we have different levels of blending,
 and instrumental broadening. 

Our approach uses the dependence of the minimisation factor $S$ on three input 
parameters, i.e. log N(Fe), \Vt and \vsini. We show that our procedure allows us to 
find only one single solution across a range of reliable parameters.
The metallicity of HD1835 is larger than the Sun, whereas HD10700 is found to 
be a metal deficient star.
In all cases we obtained realistic results using our procedure.
This provides some evidence to support the use of our procedure for studying 
the physical 
characteristics of solar-type stars.

 As we know, previous authors have fit  
 observed spectra to synthetic spectra by least squares methods to obtain  
 stellar parameters (see  Jones \& Pavlenko 2002;
 Valenti \& Fischer (2005); Jenkins et al. 2008 and references therein).
Still our procedure determines abundances, \Vt, \vsini in the 
framework of a self-consistent approach. Namely our abundances from the computed model
 atmospheres agree with the results of determinations using fits to the observed 
 spectra. We used two independent proofs of the adopted model parameters:
 a) agreement of abundances of iron obtained from the fits to Fe I and Fe II 
 lines, and b) the absence of any dependency of the iron abundance on excitation potential 
 of Fe I and Fe II lines.

In the former papers, one
or more parameters were fixed to simplify the procedure. 
Valenti \& Fisher (2005)
noted the degeneration of the procedure in respect to the \Vt determination, and 
Jenkins et al. (2008) used a fixed value of \Vt also.
 Furthermore: \\
-- Our parameters in this paper were obtained by simple averaging across 1$\sigma$ 
error space. The final results we find agree well with literature values. \\
-- We show that our solution is comparatively stable in respect to the 
small variations of 
the main input parameters: \tef, log g and [Fe/H]. Still, we carried out our 
analysis in the framework of a self-consistent approach 
where the model atmosphere agree 
with the input parameters. \\

In our analyses we used Valenti \& Fisher (2005) \tef and log g as a 
''zero approach'' value. 
We extended the standard 
procedure by using the determination of the best agreement between the results, i.e. 
log N(Fe),  
log g, \vsini and \Vt,  
obtained from the fits to Fe I and Fe II lines in the framework of 
the self-consistent model. In general, our values of \tef and log g agree well with 
Valenti \& Fisher values, but they were obtained with a smaller number of free
parameters.

In our approach we determined [Fe/H],
microturbulent velocity and \vsini from the best fits to the synthetic spectra,
where these parameters are not free. Moreover, using abundances obtained 
by fits to lines of atoms and ions in the observed spectrum, we can, 
in principle, determine \tef for these stars.
Still, abundances are 
determined here from the fits to lines of atoms and ions separately, so 
that option is a part of our procedure for other stars. Abundances 
obtained for lines of Fe I and Fe II agree well for all three of the stars
considered here. This confirms the effectiveness of our procedure. 
We note that, generally, the determination of log g is difficult. 
Still most G-dwarfs lie on the main sequence, so their log g should 
be approximately the same, around 4.5. 
Nevertheless, a more accurate log g can be determined from the 
known parallaxes. Our results show some dependence on log g. The fine analysis 
results of Tables 5 and 6 shows that we can determine all parameters from the 
best fits to Fe I and Fe II lines.

We plan to 
develop some methods to restrict possible log g values using 
fits to strong lines presented in our spectra. Wings of these lines 
are pressure broadened, so they are sensitive to log g by definition. Also 
photometric criteria can be incorporated, such that we can iterate our solutions 
using each stars position on the HR-diagram.

We carry out our computations in the framework of the LTE approach. 
In principle, the 
effects of NLTE should be accounted for. However, the realisation of the NLTE 
concept is very difficult, even for the case Fe I and Fe II, (see Mashonkina 
et al. 2011 and references therein). We cannot expect large corrections 
of our results due to NLTE effects. Refined NLTE analysis by 
Mashonkina et al. shows that the average NLTE corrections amount to changes in 
the iron abundance of 0.01 - 0.1 dex. The value is within our formal error bars
defined by our procedure and input data accuracy. The comparison of our
iron abundances in the atmospheres of the Sun and HD10700 agree well with 
Mashonkina et al. (2011) to within $\pm$0.1 dex accuracy.  
 
In general, the confidence 
of our solutions depend on the choice of input parameters. We confirm our 
solutions by comparing different parameters obtained from the fits of 
selected lines of Fe I, Fe II, and other elements to the observed spectrum. 
In fact, our results agree well with other recently published works 
for our three test cases.

Our procedure allows us to account for blending provided by other lines. 
In the most 
general case the efficiency of blending effects depend on resolution, metallicity
and rotational velocity. Our experience shows that our procedure works for 
metal deficient and metal rich stars, stars with low and intermediate low
(up to 20 km/s) rotational velocities, and over a range of \tef's. In 
the case of stars of low metallicities
or fast rotators, the number of spectral lines which can be used is 
significantly reduced.
That provides additional problems of the numerical analysis 
(see Ivanyuk et al. 2012). However, these
problems are common for all known procedures.  

We note two big advantages of our procedure: a) we use fits
of theoretically computed profiles to the features in stellar observed 
spectra. This 
provides the possibility to determine the rotational velocities from 
the fits to all lines from our lists. 
b) Our procedure can be used for the abundance  
analysis of spectral data of the same quality
obtained on high resolution spectrographs (e.g. FEROS, HARPS) for large 
numbers of stars.

\section{Acknowledgements}

The computations were carried out on the computer cluster of University of 
Hertfordshire.
Work of YP and YL was
partially supported by the ''Cosmomicrophysics'' program of the Academy of 
Sciences of
Ukraine. YP, JHJ, DP, OI acknowledge funding by
EU PF7 Marie Curie Initial Training Networks (ITN) ROPACS project (GA
N 213646).
JSJ acknowledges funding by Fondecyt through grant 3110004, along with 
partial support from Centro de Astrof\`\i sica FONDAP 15010003, the  
GEMINI-CONICYT FUND and from joint committe ESO-Government of
Chile grant.
Authors thank Nick Malygin, Joana Gomes, Bogdan Kaminsky and our three 
anonymous referees for their helpful comments.

\label{lastpage}
\end{document}